\documentclass[11pt,onecolumn,draftcls]{IEEEtran}
\include{func}
\usepackage{amsmath}
\usepackage{enumerate}
\usepackage{rotating}
\usepackage{amsfonts}
\usepackage{amssymb}
\usepackage{psfrag}
\usepackage{psfrag}
\usepackage{cite,color,url,path,multirow,multicol}
\usepackage{subfigure}
\usepackage{cite,color,url}
\usepackage{subfigure}
\usepackage{epsfig}
\usepackage{yhmath}
\usepackage{algorithm}
\usepackage{algorithmic}

\newcommand{\smatlabaxislabel}[1]{\fontsize{12}{\f@baselineskip}%
\textsf{#1}}
\newcommand{\matlabaxislabel}[1]{\fontsize{14.4}{\f@baselineskip}%
\textsf{#1}}
\newcommand{\mmatlabaxislabel}[1]{\fontsize{17.28}{\f@baselineskip}%
\textsf{#1}}
\newcommand{\bmatlabaxislabel}[1]{\fontsize{20.74}{\f@baselineskip}%
\textsf{#1}}
\newcommand{\bbmatlabaxislabel}[1]{\fontsize{24.88}{\f@baselineskip}%
\textsf{#1}} \makeatother

\allowdisplaybreaks

\newcommand{\beq}{\begin{equation}}
\newcommand{\eeq}{\end{equation}}

\newcommand{\nn}{\nonumber}

\title{ \textbf{ Statistical Simulation Models for Cascaded Rayleigh Fading Channels}}
\author{{\normalsize Yazan Ibdah and Yanwu Ding}
\thanks{The authors are with the Department of Electrical Engineering and Computer
Science, Wichita State University, 1845 N. Fairmount, Wichita, KS 67260.  Emails : \{yxibdah,  yanwu.ding\}@wichita.edu.
Part of the material in this paper was presented at the IEEE Military Communications Conference, MILCOM 2011 in Baltimore, Maryland, USA, November 2011.}}

\begin{document}
\maketitle

\vspace{-2.25cm}
\begin{abstract}
\vspace{-0.5cm}

 In this paper, we  present statistical simulators  for  cascaded Rayleigh fading channels with and without line-of-sight~(LOS).  These simulators contain two individual summations and are therefore easy to implement with  lower complexity.  Detailed  statistical properties, including auto- and cross-correlations of the in-phase, quadrature components of the channels, envelopes, and squared envelopes, are  derived. The time-average statistical properties and the corresponding variance are also investigated  to justify that the proposed simulators achieve good convergence performance.
 Extensive Monte Carlo simulations are performed for various statistical properties to validate the proposed simulators. Results show that the simulators provide fast convergence  to all desired statistical properties, including the probability density function~(PDF), various auto- and cross-correlations, level crossing rate~(LCR), and average fading duration~(AFD).
 While various tests and measurements in dense scattering urban and forest environments indicate that mobile-to-mobile channels may experience cascaded Rayleigh fading, the proposed statistical  models can be applied to simulate  the underlying channels.


\end{abstract}

\vspace{0.2cm}
\vspace{0.1cm}

\section{Introduction} \label{sec_intro}

Mobile-to-mobile communications have found more and more applications in systems where the terminals and nodes are no longer stationary~\cite{R7, R8}, such as mobile ad hoc wireless networks~\cite{R1}, wireless local area networks~\cite{R2}, intelligence transportation systems~\cite{R3, R4, R5}, and vehicular-to-vehicular systems~\cite{R6}. Consequently, successful development of statistical models or simulators  for mobile-to-mobile~(M-M) channels is inevitable  in order to understand the  statistical properties of the underlying channels and effective designs for the systems.
A pioneering  mathematical reference model for M-M channels is found in the earlier work of Akki and Haber~\cite{R14} and Akki~\cite{R15}. This analysis is  then extended to account for scattering in three dimensions by Vatalaro and Forcella~\cite{vatalaro_vt_1997}.
  A modified  statistical model for suburban outdoor-to-indoor
M-M communications channels has been discussed in~\cite{R25}. A double-summation-of-sinusoids statistical model is applied to simulate M-M channels, assuming omni-directional antennas and isotropic scattering  around the transmitter (Tx) and the receiver (Rx)~\cite{R16}. Due to the uniform scattering around the Tx and Rx, the  statistical model in~\cite{R16} is also referred to as a ``double-ring" statistical model~(the single-ring statistical model is proposed for fixed-to-mobile channels~\cite{R18}).
The double-ring model is further modified to generate multiple uncorrelated complex faded envelopes using orthogonal in-phase and quadrature components~\cite{R17}. The modification offers a faster convergence and lower variance at higher  complexity.  The double-ring concept is also applied to modeling M-M channels where a line-of-sight  component exists between the Tx and Rx~\cite{R19}~(the single-ring-based fixed-to-mobile Rican channel models can be found in~\cite{R20, R21}).
In~\cite{R31}, stochastic properties for M-M narrow-band  channels are derived from a geometrical two-ring scattering model, assuming that both the transmitter and the receiver are surrounded by an infinite number of local scatters.  The model is further extended to a three-ring scattering model~\cite{goemetical_channel_vtc_2009}.

While the double-ring model characterizes a category of mobile-to-mobile channels that follow a Rayleigh fading distribution, various measurements in dense scattering urban and forest environments where signals diffract from street corners, building edges, and moving vehicles, suggest that if large separations between the Tx and Rx are much larger than both radii rings~\cite{R31} at the TX and Rx, then the mobile-mobile channels follow fading distributions that are more severe than a single Rayleigh fading.
 In~\cite{R25}, the outdoor-to-indoor M-M radio channel measurements show that the received power follows a double-Rayleigh distribution rather  than a single Rayleigh distribution. Similar observations are seen in the measurements~\cite{R27}, whereas the Tx is moving in within a circle of a 1 m and the RX are  randomly chosen within a 10 m by 10 m square with no line of sight.  The measurements in~\cite{R43,R23,R26} indicate that severe (worse than Rayleigh) fading in vehicle-vehicle channels  is observed in small cities, urban, and suburban areas, motorways, and highways.

In~\cite{R28, R29}, the concept of a mixture of single, double, and triple Rayleigh distributions is introduced to represent a general M-M channel. The general multiple Rayleigh channel transfer function was initially suggested in~\cite{R25} for studying dense scattering environments. Effective statistical simulators for  cascaded or multiple Rayleigh fading M-M channels seem sparse in the literature.  
 In this paper, we develop two statistical simulators for a cascaded Rayleigh fading channel with no line-of-sight~(NLOS)  and extend the models to channels with LOS.  The simulators contain two individual summations; therefore, they are easy to implement with  lower complexity. With appropriately chosen parameters, the proposed models can achieve the desirable statistical  properties in a small number of trials, and provide satisfactory convergence performance.   Detailed  statistical properties, including auto- and cross-correlations of the in-phase, quadrature components of the channels, envelopes, and squared envelopes are  derived for the simulators. The time-average statistical properties and  corresponding variance are also investigated  to justify that the proposed simulators can achieve  good convergence performance.
  Extensive Monte Carlo simulation results are provided for various statistical properties to validate the proposed simulators.



The remainder of this paper is organized as follows: Section II provides a brief review of the general multiple Rayleigh channel transfer function and presents the proposed channel models without LOS and with LOS. Section III presents detailed statistical properties of the proposed models.  Simulations results are provided in Section IV, and Section V concludes the paper with a summary of observations.

\section{Statistical Channel Simulators for Cascaded Rayleigh Fading Channels}

\subsection{Transfer function of cascaded Rayleigh fading channels}
In highly dense scattering environments,  the transmission path between the Tx and Rx are composed of many  scattering paths. Fig.~\ref{F1} illustrates a multiple scattering environment between two mobile terminals with no LOS exists between the Tx and Rx (first graph) and a LOS (second graph).
 We assume that the distance between the transmitter and receiver is much larger than  both radii of the scattering rings at the Tx and Rx,  the channel in the first graph follows  cascaded Rayleigh distribution with NLOS, and that in the second graph cascaded Rayleigh distribution with LOS.

A cascaded Rayleigh transfer function is suggested in~\cite{R25}, assuming that the two groups of scatter at the transmitter and the receiver are stationary and located a relatively large distance apart with a separation distance of $D\gg R_t+R_r$, where $R_t$ and $R_r$ are the radii of the scatters around the Tx and Rx, respectively. The transfer function is presented as
\begin{align}
g_{\text{general}}(t)=\sum_{n=1}^{N} {A}_n {G}_{\small_{T}} (k_{{\small_{T}}_n})e^{j(\omega_{{\small_{T}}_n}t+\Phi_{{\small_{T}}_n})}
\sum_{m=1}^{M} {B}_m {G}_{\small_{R}} (k_{{\small_{R}}_m})e^{j(\omega_{{\small_{R}}_m}t+\Phi_{{\small_{R}}_m})} \label{y1}
\end{align}
where subscripts T and R are associated with Tx and Rx, respectively, ${A}_n,   \Phi_{{\small_{T}}_n}, n=1,\cdots, N$, are
the identically distributed random (i.i.d) amplitude and phase of $N$ scatter components around the Tx, $\ {B}_m, \Phi_{{\small_{R}}_m}, m=1,\cdots, M$ are the i.i.d amplitude and phase of $M$  around the Rx, ${G}_{\small_{T}}$ and ${G}_{\small_{R}}$ are antennas gains, $k_{{\small_{T}}_n}$ and $k_{{\small_{R}}_m}$ are unit vectors corresponding, respectively, to the direction of departure and arrival of the $n$-th scatter at Tx and $m$-th scatter at the Rx, and $\omega_{{\small_{T}}_n}$ and $\omega_{{\small_{R}}_m}$ are Doppler spreads of the Tx and Rx.  Efficient statistical simulators for~\eqref{y1} were left unexplored. In this section, we develop four channel simulators~(Simulators A, B, C, and D) for the cascaded Rayleigh fading channels without and with LOS.
%
%

\subsection{Statistical channel simulators without LOS}
In this subsection, two simulators, namely Simulators A and B, are presented for cascaded Rayleigh fading without LOS in~\eqref{y1}.

\subsubsection{Simulator A} In the statistical channel model, the  complex scattering components in cascaded Rayleigh fading channels are given by
\begin{align}
g_{\small_A}(t)=\sqrt{\frac{\sqrt{2}}{Q}} \sum_{n=1}^{Q} e^{j\big(2 \pi f_1 t \cos(\gamma_n)+\theta_n\big)}
\sqrt{\frac{\sqrt{2}}{P}} \sum_{m=1}^{P} e^{j\big(2 \pi f_2 t \cos(\zeta_m)+\Phi_m\big)} \label{y2}
\end{align}
where $Q$ and  $P$ designate the number of scatters around the Tx and Rx, respectively, $f_1$ and $f_2$ are the Doppler's shift frequencies,
$\theta_n$ and $\Phi_m\in [-\pi, \pi), n=1, \cdots, Q, m=1, \cdots, P $, are the phase shifts for each scatter from the Tx and to the Rx and they are i.i.d for all $n$ and $m$. $\gamma_n$  represents the angle of departure for the  $n$-th scatter at the Tx, and $\zeta_m$ is the angle of arrival for the $m$-th  scatter at the Rx, given by, respectively,
\begin{align}
\gamma_n=\frac{2n \pi -\pi+\psi}{4Q }, \ \zeta_m=\frac{2m \pi -\pi+\varphi}{2P} \label{y4},
\end{align}
where $\psi, \ \varphi$ are independent and uniformly distributed in $ [-\pi, \pi)$. While those angles can be chosen as either dependent on or independent of $n$ and $m$~\cite{R19, R21,R16, R18}, we choose them to be independent in order to reduce the complexity of the simulator.

\subsubsection{Simulator B} The  angles of departures and arrivals for Simulator A, as indicated in~\eqref{y4}, are in $[0, \pi/2)$ and $[0, \pi)$, respectively. In fact, smaller angles of departures and arrivals in the scatters  are expected for the channels under discussion,  because  the distance between  Tx and Rx is relatively large with  $D \gg  R_t+R_r$. However, our simulations suggest that if the range of angles of arrivals is chosen smaller, the convergence performance of Simulator A degrades, especially for the cross-correlation between the in-phase and quadrature components. We seek an alternative simulator with an improved performance and  the angles of departures and arrivals are both in $[0, \pi/2)$ to achieve a better representation of scattering patterns in the channels between a TX and Rx with larger separations.

%

We assume that Simulator B has the following form: $
 g_{\small_B}(t) =\big(g_{1c}(t) +j g_{1s}(t)\big) \big(g_{2c}(t) +j g_{2s}(t)\big)$,
where
$ g_{ic}(t)$ and $ g_{is}(t), i=1, 2, $  are the sinusoidal functions which characterize, respectively, the scattering at the Tx and Rx. Here, we present the steps to obtain the functions at Tx, $ g_{1c}(t)$ and $g_{1s}(t)$, while same procedure applies to obtain  $ g_{2c}(t)$ and $g_{2s}(t)$ at the Rx.

Let $\alpha_k=\frac{2k \pi -\pi+\psi}{K_{\text{Tx}}}$, and $\theta_k,~ k=1, \cdots, K_{\text{Tx}}, $ denote  the  angle of departure and phase for the $k$-th scatter at the Tx, where  $ \psi,~ \theta_k $ are statistically independent and uniformly distributed in $ [-\pi, \pi)$ for all $k$, and $ K_{\text{Tx}}$ is the number of scatters around the Tx.
Next, we evaluate a summation of series of exponentials:
$ \sum_{k=1}^{K_{\text{Tx}}} \exp\big(j(2\pi f_1 t \cos(\alpha_k)+\theta_k)\big)$. While the summation admits a similar expression as the first summation in Simulator A, we consider a special case, $K_{\text{Tx}}=4N$ with  $N>1$ being an integer.
The summation  is obtained as four terms corresponding to the quarters of $K_{\text{Tx}}$,
 $ \sum_{k=1}^{K_{\text{Tx}}} \exp\big(j(2\pi f_1 t \cos(\alpha_k)+\theta_k)\big)=\sum_{\ell=0}^3 \Omega_\ell$, where
$ \Omega_\ell= \sum_{n=\ell N+1}^{(\ell+1)N} \exp\Big(j\big(2\pi f_1 t \cos(\alpha_n)+\theta_n\big)\Big)= \sum_{n=1}^{N} \exp\Big(j\big(2\pi f_1 t \cos(\alpha_n+\frac{\ell \pi}{2})+\theta_{n+\ell N}\big)\Big),~ \ell=0, \cdots,3$.
Introducing  uniform distributed random phases $\Theta_n \in [-\pi, \pi), n=1, \cdots, N$, where $\Theta_n$ is independent of $\psi$, $\theta_k$ for all $n$ and $k$, one can rewrite $ \Omega_i, i=1, 2, 3 $  as
\begin{align}
\Omega_1&=
\sum_{n=1}^{N} \exp\Big(j\big(2\pi f_1 t \cos(\alpha_n+\frac{\pi}{2})-\Theta_n \big)\Big), \label{Gamma_1} \\
\Omega_2&=
\sum_{n=1}^{N} \exp\Big(j\big(2\pi f_1 t \cos(\alpha_n+\pi )-\theta_{n}\big)\Big), \label{Gamma_2} \\
\Omega_3&=  \sum_{n=1}^{N} \exp\Big(j\big(2\pi f_1 t \cos(\alpha_n+\frac{3\pi}{2} )+\Theta_n \big)\Big). \label{Gamma_3}
\end{align}
The range of angles of departure in $\Omega_\ell,~ \ell=0,\cdots, 3 $ is in $(0, \pi/2)$ as specified by $ \alpha_n,~ n=1, \cdots, N $~(the range give by $ \alpha_k,~ k=1, \cdots, K_{\text{Tx}} $ is in $(0, \pi/2)$).
 We assign the sinusoidal functions at the Tx, i.e. $g_{1c}(t)$ and $g_{1s}(t)$  equal, respectively, to the terms associated with $\cos(\alpha_n)$ and $\sin(\alpha_n)$ in $\sum_{\ell=0}^3 \Omega_\ell$, along with a normalizing factor.   Notice $\Omega_0+\Omega_2= 2\sum_{n=1}^{N} \cos\big(2\pi f_1 t \cos(\alpha_n)+\theta_{n}\big) $, and $\Omega_1+\Omega_3= 2\sum_{n=1}^{N} \cos\big(2\pi f_1 t \sin(\alpha_n)+\Theta_{n}\big) $.   We obtain $ g_{1c}(t)=\sqrt{\frac{\sqrt{2}}{N}}
 \sum_{n=1}^{N} \cos\big(2\pi f_1 t \cos(\alpha_n)+\theta_{n}\big) $, and  $ g_{1s}(t)=\sqrt{\frac{\sqrt{2}}{N}}
 \sum_{n=1}^{N} \cos\big(2\pi f_1 t \sin(\alpha_n)+\Theta_{n}\big) $ for Simulator B. It is worth to mention that the choice of $g_{1c}(t)$ and $g_{1s}(t)$ at the Tx coincide with the terms in the  statistical model of fixed-to-mobile cellular channels~\cite{R18}.

 Let  $ \beta_k=\frac{2k \pi -\pi+\varphi}{K_{\text{Rx}}}$, and $ \Phi_k,~k=1,\cdots, K_{\text{Rx}} $
     denote the angle of arrival and phase for the $k$-th scatter  at the Rx, where  $ \varphi,~ \Phi_k $ are uniformly distributed in $ [-\pi, \pi)$, and they are  impendent for all $k$, and $ K_{\text{Rx}}$ is the number of scatters around the Rx.  Evaluating the summation $ \sum_{k=1}^{K_{\text{Rx}}} \exp\Big(j\big(2\pi f_2 t \cos(\beta_k)+\Phi_k\big)\Big)$ for $K_{\text{Rx}}=4M$, and following similar procedure describe above, we obtain $g_{2c}(t)=\sqrt{\frac{\sqrt{2}}{M}}
 \sum_{m=1}^{M} \cos\big(2\pi f_2 t \cos(\beta_m)+\Phi_m\big) $, and  $g_{2s}(t)=\sqrt{\frac{\sqrt{2}}{M}}
 \sum_{m=1}^{M}  \cos \big(2\pi f_2 t \sin(\beta_m)+\Psi_m\big) $ for Simulator B. The range of angles of arrivals at the Rx is in $(0,  \pi/2)$ as specified  by $\beta_m,~ m=1, \cdots, M $.
In summary,  Simulator B is expressed as
\begin{align}
g_{\small_B}(t) =\big(g_{1c}(t) +j g_{1s}(t)\big) \big(g_{2c}(t) +j g_{2s}(t)\big)  \label{y5}
\end{align}
where
$ g_{ic}(t)$ and $g_{is}(t), i=1, 2 $ are defined as
\begin{align}
g_{1c}(t)&=\sqrt{\frac{\sqrt{2}}{N}} \sum_{n=1}^{N}  A_n(t), g_{1s}(t)=\sqrt{\frac{\sqrt{2}}{N}} \sum_{n=1}^{N}  C_n(t), \label{y8}\\
g_{2c}(t)&=\sqrt{\frac{\sqrt{2}}{M}} \sum_{m=1}^{M}  B_m(t),
g_{2s}(t)=\sqrt{\frac{\sqrt{2}}{M}} \sum_{m=1}^{M}  D_m(t), \label{y10}
\end{align}
$A_n(t)=\cos\big(2\pi f_1 t \cos(\alpha_n)+\theta_n\big),
C_n(t)=\cos \big(2\pi f_1 t \sin(\alpha_n)+\Theta_n\big), n=1, \cdots, N, B_m(t)=\cos \big(2\pi f_2 t \cos(\beta_m)+\Phi_m\big)$,  and $ D_m(t)=\cos \big(2\pi f_2 t \sin(\beta_m)+\Psi_m\big), m=1, \cdots, M$,
$f_1$ and $f_2$ are the Doppler's shift frequencies, $\theta_n, \ \Theta_n\in [-\pi, \pi), n=1, \cdots, N$  are the phase shifts in the $n$-scatter at the Tx, $\Phi_m, \ \Psi_m, \in [-\pi, \pi), m=1,\cdots, M$  are the phase shifts in the $m$-scatter at the Rx and are i.i.d  for all $n$ and $m$,  $\alpha_n$  is the angle of departure of the $n$-th scatter at the Tx, and $\beta_m $ is the angle of arrival of the $m$-th scatter at the Rx, and they are calculated, respectively, as
\begin{align}
\alpha_n=\frac{2n \pi -\pi+\psi}{4N}, \ \beta_m=\frac{2m \pi -\pi+\varphi}{4M},  \label{z2}
\end{align}
where $ \psi, \ \varphi$ are  independent and uniformly distributed in $ [-\pi, \pi)$, and they are independent to all the phases at the Tx and Rx.
 It is indicated  that a statistical M-M model may result in faster convergence rates by choosing a smaller range of angles of arrivals and departures~\cite{R16}.  Our simulations justify the observation as well.
 Compared with Simulator A,  Simulator B has faster convergence and requires fewer number of trials to converge to the desired statistical properties.



\subsection{Statistical channel simulators with LOS}


Adding the LOS component to Simulators A and B, we present two simulators for cascaded Rayleigh fading channels with LOS, namely Simulators C and  D, as
%
\begin{align}
&\text{Simulator C: } h_{\small_C}(t)=\frac{g_{\small_A}(t) + \sqrt{2K} e^{j(2 \pi f_3 t \cos(\phi_3)+\phi_0 )}}{\sqrt{2(1+K)}}, \label{y11}\\
&\text{Simulator D: } h_{\small_D}(t)=\frac{g_{\small_B}(t) + \sqrt{2K} e^{j(2 \pi f_3 t \cos(\phi_3) +\phi_0)}}{\sqrt{2(1+K)}}, \label{y14}
\end{align}
where $K$ is spectral to the scatter power ratio, $\phi_0$  is uniformly distributed in $[-\pi, \pi)$, $f_3$ is the Doppler frequency caused by the relative velocity, because  both Tx and Rx have mobility, $\phi_3$ is the relative angle
between the relative movement and the LOS component, and the values of $f_3$ and $ \phi_3$ are given, respectively, by~\cite{R19} as 
\begin{align}
f_3&=\frac{\Big|\sqrt{\big(|v_1|\cos(\phi_{12}) -|v_2|\big)^2+\big( |v_1|\sin(\phi_{12})\big)^2} \Big|}{\lambda} \label{y12}\\
\phi_3&=\cos^{-1}\biggl(\frac{|v_1|^2+|v_3|^2-|v_{2}|^2}{2|v_1||v_3| } \biggl)+\phi_1 \label{y13}
\end{align}
where $v_1$ and $v_2$ are, respectively, the speeds of the Tx and Rx,  $v_3$ is the relative speed calculated by $v_3= f_3 \lambda$,   $\lambda$ is the wavelength of the carrier,  $\phi_1$ is the angle between the Tx and the LOS, and $\phi_{12}$ is the angle between the Tx and Rx directions.
%
%
%
%
Simulator D has a slightly additional complexity  compared to Simulator C, due to the higher complexity in $g_B(t)$ than in $g_A(t)$. However,    as indicated by the simulation results, Simulator D provides faster convergence to the desired statistical properties  even for a lower value of $K$.

\section{Statistical Properties of Proposed Models}

 Because Simulators A and B, $g_{\small_{A}} (t)$ and $g_{\small_{B}} (t)$,   have the same statistical properties, as do Simulators C and D, $h_{\small_{C}} (t)$ and $h_{\small_{D}} (t)$,  we focus on Simulators B and  D when representing the statistical properties in this section. To make notations less bulky, we drop the subscripts, and  use $g(t)$ and $h(t)$ to present, respectively, the simulators without and with LOS.

\subsection{Second-order statistics for Simulator B}
Let  $g_c (t)=\text{Re}\big(g(t)\big)$ and $ g_s (t)=\text{Im}\big(g(t)\big)$ be the real~(in-phase) and imaginary~(quadrature) parts of Simulator B.
The  autocorrelation, the cross-correlation of the in-phase, quadrature components, and the autocorrelation of complex envelopes are given below. Steps for the proof are presented in Appendix~\ref{A7}.
\begin{align}
R_{g_cg_c}(\tau)&=\mathbb{E}\big[g_c(t+\tau)g_c(t) \big]=J_0(2\pi f_1 \tau)J_0(2\pi f_2\tau)  \label{y21}\\
R_{g_sg_s}(\tau)&=\mathbb{E}\big[g_s(t+\tau)g_s(t) \big]=R_{g_cg_c}(\tau) \label{y24}\\
R_{g_cg_s}(\tau)&=\mathbb{E}\big[g_c(t+\tau)g_s(t) \big]=0, R_{g_sg_c}(\tau)=\mathbb{E}\big[g_s(t+\tau)g_c(t) \big]=0 \label{y22}\\
R_{gg}(\tau)&=\frac{1}{2}\mathbb{E}\big[g(t+\tau) g^*(t) \big] \nn \\
&=\frac{1}{2}\big(R_{g_cg_c}(\tau)+R_{g_cg_s}(\tau)+R_{g_sg_c}(\tau)+R_{g_sg_s}(\tau) \big)=J_0(2\pi f_1 \tau)J_0(2\pi f_2\tau) \label{y26}
\end{align}
\begin{align}
&~~~~~R_{|g|^2|g|^2}(\tau)=\mathbb{E}\big[g_c^2(t)g_c^2(t+\tau) +g_s^2(t)g_c^2(t+\tau) +g_c^2(t)s_s^2(t+\tau) +g_s^2(t)g_s^2(t+\tau) \big]\nn\\
&~~~=4+4J_0^2(2\pi f_1 \tau)+4J_0^2(2\pi f_2 \tau)+4J_0^2(2\pi f_1 \tau)J_0^2(2\pi f_2 \tau)+\frac{J_0(4\pi f_1 \tau)J_0(4\pi f_2 \tau)}{4NM}\nn\\
&~~~+\frac{J_0(4\pi f_1 \tau)+J_0(4\pi f_1 \tau)J_0^2(2\pi f_2 \tau)}{N}+\frac{J_0(4\pi f_2 \tau)+J_0(4\pi f_2 \tau)J_0^2(2\pi f_1 \tau)}{M}+\frac{4 \xi(f_1,\tau)\xi(f_2,\tau)}{N^2M^2} \nn\\
&~~~-\frac{4M\big( 1+ J_0^2(2\pi f_2 \tau)\big)+J_0(4\pi f_2\tau)}{N^2M}\xi(f_1,\tau)
-\frac{4N\big( 1+ J_0^2(2\pi f_1 \tau)\big)+J_0(4\pi f_1\tau)}{NM^2}\xi(f_2,\tau)
\label{ynew32}
\end{align}
where $J_0$ is the zero-order Bessel function  first kind, $ \xi(f_1,\tau)= \sum_{n=1}^{N} \Big( \mathbb{E}\big[\cos\big(2\pi f_1 \tau \cos(\alpha_n) \big) \big]\Big)^2$, and $\xi(f_2,\tau)=\sum_{m=1}^{M} \Big( \mathbb{E}\big[\cos\big(2\pi f_2 \tau \cos(\beta_m) \big) \big]\Big)^2$.
The  expectation terms in~\eqref{ynew32}  can  be evaluated numerically.
It is also worth noting that, although Simulator B has a different probability density distribution function than the double-ring simulator in~\cite{R16},  their  autocorrelation and cross-correlation properties are the same.

The time-average correlations for Simulator B can be derived as
\begin{align}
\tilde{R}_{g_c g_c}(\tau)&=\lim_{T\rightarrow \infty} \frac{1}{T} \int_{0}^{T} g_c(t) g_c(t+\tau) dt = \frac{1}{2MN }
\sum_{n,m=1}^{N,M} \big(\tilde{A}_n(\tau) \tilde{B}_m(\tau) +\tilde{C}_n(\tau) \tilde{D}_m(\tau)\big) \\ \label{n1}
\tilde{R}_{g_c g_s}(\tau)&=\lim_{T\rightarrow \infty} \frac{1}{T} \int_{0}^{T} g_c(t) g_s(t+\tau) dt = 0 \\ \label{n2}
\tilde{R}_{g_s g_c}(\tau)&=\lim_{T\rightarrow \infty} \frac{1}{T} \int_{0}^{T} g_s(t) g_c(t+\tau) dt = 0 \\ \label{n3}
\tilde{R}_{g_s g_s}(\tau)&=\lim_{T\rightarrow \infty} \frac{1}{T} \int_{0}^{T} g_s(t) g_s(t+\tau) dt = \frac{1}{2MN }
\sum_{n,m=1}^{N,M} \big(\tilde{A}_n(\tau) \tilde{D}_m(\tau) +\tilde{C}_n(\tau) \tilde{B}_m(\tau)\big)  
\end{align}
where $\tilde{A}_n(\tau)\!=\!\cos\big(2\pi f_1 \tau \cos(\alpha_n) \big)$,  $\tilde{B}_m(\tau)\!=\!\cos\big(2\pi f_2 \tau \cos(\beta_m) \big)$, $ \tilde{C}_n(\tau)\!=\!\cos\big(2\pi f_1 \tau \sin(\alpha_n) \big)$, and $\tilde{D}_m(\tau)=\cos\big(2\pi f_2 \tau \sin(\beta_m) \big)$.
The variance of the time-average correlations for Simulator B can be derived as
\begin{align}
&\text{Var}[\tilde{R}_{g_c g_s} (\tau)] =\mathbb{E}[ \tilde{R}_{g_c g_s}^2(\tau)]-({R}_{g_c g_s}(\tau))^2=0 \label{r1}\\
&\text{Var}[\tilde{R}_{g_s g_c} (\tau)] =\mathbb{E}[ \tilde{R}_{g_s g_c}^2(\tau)]-({R}_{g_s g_c}(\tau))^2=0\label{r2}\\
&\text{Var}[\tilde{R}_{g_s g_s}(\tau) ] =\mathbb{E}[ \tilde{R}_{g_s g_s}^2(\tau)]-({R}_{g_s g_s}(\tau))^2=\text{Var}[\tilde{R}_{g_c g_c}(\tau)] \label{r3}\\
&\text{Var}[\tilde{R}_{g_c g_c}(\tau) ] =\mathbb{E}[ \tilde{R}_{g_c g_c}^2(\tau)]-({R}_{g_c g_c}(\tau))^2\nn\\
&=\frac{(1\!+\!J_0(4 \pi f_1 \tau)) (1\!+\!J_0(4\pi f_2 \tau))\!+\!2N J_0^2(2\pi f_1\tau)(J_0(4\pi f_2\tau)\!+\!1)\!+\!2MJ_0^2(2\pi f_2\tau)(J_0(4\pi f_1\tau)\!+\!1)}{8NM}\nn\\
&+\frac{J_0^2(2 \pi f_2 \tau)}{2N^2}V_{\tilde{A}\tilde{C}}
+\!\frac{J_0^2(2 \pi f_1 \tau)}{2M^2}V_{\tilde{B}\tilde{D}}
+\frac{1}{2N^2 M^2}\Big( V_{\tilde{A}\tilde{C}}V_{\tilde{B}\tilde{D}}+\xi(f_1,\tau)\xi(f_2,\tau)\Big)\nn\\
&-\frac{1\!+\!J_0(4\pi f_2 \tau)\!+\! 2MJ_0^2(2\pi f_2 \tau)}{4N^2M}\xi(f_1,\tau)
-\frac{1\!+\!J_0(4\pi f_1 \tau)\!+\! 2NJ_0^2(2\pi f_1 \tau)}{4NM^2}\xi(f_2,\tau)
\label{r4}
\end{align}
where
$V_{\tilde{A}\tilde{C}}= \sum_{n=1}^N \big( \mathbb{E}[ \tilde{A}_n(\tau) \tilde{C}_n(\tau)]-\mathbb{E}[ \tilde{A}_n(\tau)]\mathbb{E}[ \tilde{C}_n(\tau)] \big)$ and
$V_{\tilde{B}\tilde{D}}= \sum_{m=1}^M \big( \mathbb{E}[ \tilde{B}_m(\tau) \tilde{D}_m(\tau)]-\mathbb{E}[ \tilde{B}_m(\tau)]\mathbb{E}[ \tilde{D}_m(\tau)] \big)$.
\subsection{ Second-order statistics for Simulator D}

Autocorrelation, cross-correlation, autocorrelation of the complex envelopes, and autocorrelation of the squared envelope for Simulator D are given below. Steps for the proof are presented in Appendix~\ref{A3}.
\begin{align}
R_{h_ch_c}(\tau)&=\mathbb{E}\big[h_c(t+\tau)h_c(t) \big]=\frac{J_0(2\pi f_1 \tau)J_0(2\pi f_2\tau)+K\cos\big(2\pi f_3\tau \cos(\phi_3) \big)}{2(1+K)} \label{y27}\\
R_{h_ch_s}(\tau)&=\mathbb{E}\big[h_c(t+\tau)h_s(t) \big]=-\frac{K\sin\big( 2\pi f_3 \tau \cos(\phi_3)\big)}{2(1+K)}\label{y29}\\
R_{h_sh_s}(\tau)&=\mathbb{E}\big[h_s(t+\tau)h_s(t) \big]=R_{h_ch_c}(\tau) \label{y28}\\
R_{h_sh_c}(\tau)&=\mathbb{E}\big[h_s(t+\tau)h_c(t) \big]=- R_{h_ch_s}(\tau)\label{y30}\\
R_{hh}(\tau)&\!=\frac{1}{2}\mathbb{E}\big[h(t+\tau) h^*(t)\big]=\frac{1}{2}\big(R_{h_ch_c}(\tau)+R_{h_sh_s}(\tau)-jR_{h_ch_s}(\tau)+jR_{h_sh_c}(\tau) \big)\nn \\
&~~~~~~~~~~~~~~~~~~~~~~~~~~=\frac{J_0(2\pi f_1 \tau)J_0(2\pi f_2\tau)+K e^{j2\pi f_3 \tau \cos(\phi_3)}}{1+K} \label{y31}
\end{align}
\begin{align}
R_{|h|^2|h|^2}(\tau)=\mathbb{E}\big[h_c^2(t)h_c^2(\varrho) +h_s^2(t)h_c^2(\varrho) +h_c^2(t)h_s^2(\varrho) +h_s^2(t)h_s^2(\varrho) \big]~~~~~~~~~~~~~~~~~\label{correlation_square_hh}\\
=\frac{ R_{|g|^2|g|^2}(\tau)
 +8K R_{g_c g_c}(\tau) \cos\big(2\pi f_3 \tau \cos(\phi_3)\big)\!+\!{8K\!+\! 4K^2\!}}{4(1+K)^2} ~~~~\label{y32}~~~~~~~~
\end{align}

For sufficient $N$ and  $M$, 
the expression of~\eqref{y32} can be simplified as 
\begin{align}
R_{|h|^2|h|^2}(\tau)&=\frac{1}{(1+K)^2}\Big(1+J_0^2(2\pi f_1 \tau)+J_0^2(2\pi f_2 \tau)+J_0^2(2\pi f_1 \tau)J_0^2(2\pi f_2 \tau) \nn\\
&~~~~~~~~~~~~~~~~~~+2K\big(1+J_0(2\pi f_1\tau)J_0(2\pi f_2\tau) \cos(2\pi f_3\tau\cos(\phi_3) )\big)+K^2\Big).\label{y33}
\end{align}

The time-average correlations for Simulator D can be expressed as
\begin{align}
\tilde{R}_{h_c h_c}(\tau)=\frac{\tilde{R}_{g_c g_c}(\tau) +K \cos\big( 2\pi f_3 \tau \cos(\phi_3) \big)}{2(1+K)} \label{n5}\\
\tilde{R}_{h_c h_s}(\tau)= \tilde{R}_{h_s h_c}(\tau)=\frac{K\sin\big( 2 \pi f_3 \tau \cos(\phi_3)\big)}{2(1+K)} \label{n7}\\
\tilde{R}_{h_s h_s}(\tau)=\frac{\tilde{R}_{g_s g_s}(\tau) +K \cos\big( 2\pi f_3 \tau \cos(\phi_3) \big) }{2(1+K)}.  
\end{align}

The variance of the time-average correlations for Simulator D can be derived as
\begin{align}
\text{Var}[\tilde{R}_{h_c h_c} (\tau)] &=\mathbb{E}[ \tilde{R}_{h_c h_c}^2(\tau)]-({R}_{h_c h_c}(\tau))^2=\frac{\text{Var}[\tilde{R}_{g_c g_c}(\tau)]}{4(1+K)^2}  \label{r5}\\
\text{Var}[\tilde{R}_{h_s h_s} (\tau)] &=\mathbb{E}[ \tilde{R}_{h_s h_s}^2(\tau)]-({R}_{h_s h_s}(\tau))^2=\frac{\text{Var}[\tilde{R}_{g_s g_s}(\tau)]}{4(1+K)^2}  \label{r6}\\
\text{Var}[\tilde{R}_{h_c h_s} (\tau)] &=\mathbb{E}[ \tilde{R}_{h_c h_s}^2(\tau)]-({R}_{h_c h_s}(\tau))^2=0 \label{r7}\\
\text{Var}[\tilde{R}_{h_s h_c} (\tau)] &=\mathbb{E}[ \tilde{R}_{h_s h_c}^2(\tau)]-({R}_{h_s h_c}(\tau))^2=0\label{r8}
\end{align}

\section{Simulations Results}

In this section, we present extensive Monte Carlo simulation results on the statistical properties of the proposed simulators. The simulation results are obtained using $N_s=10^6$ samples in a $10$s time duration with sampling period $T_s=10^{-5} s$. The frequencies in all simulators are $f_1= f_2=100Hz$,  $P=Q=16$ in Simulators A and C,  $M=N=16$ in Simulators B and D, and $\phi_1=\phi_{12}=\frac{\pi}{2}$ in Simulators C and D. All plots are obtained using one trial unless stated otherwise.   We also simulate the double-ring~\cite{R16} and fixed-to-mobile channel models~\cite{R18, R20, R38} to demonstrate the difference in  statistical behavior of cascaded Rayleigh fading and single Rayleigh fading distributions,


\subsection{ Probability density functions for channel envelopes }


The distribution of channel envelopes for cascaded Rayleigh channels without and with LOS has been discussed in~\cite{R25, R33,R34}. Recognizing that the average power for simulators A and  B is one,  the probability density function (PDF), the cumulative density function (CDF) of the envelopes of Simulators A and  B are given, respectively, by (major steps are provided in the Appendix~\ref{A1} for convenience of readers):
 $ p_{Z}(z)=2 z K_0\big(z\sqrt{2}\big)$, and $P_{Z}(Z \leq z)=1-\sqrt{2} z K_1\big(z\sqrt{2}\big)$,  
where $z$ represents the envelope of Simulators A and  B, i.e., $ |g(t)|$,  $K_0$ and $K_1$ are, respectively, the zero-order and first-order second kind modified Bessel function.
Using (12) in~\cite{R34} with
$\sigma_i=\frac{1}{\sqrt{2\sqrt{1+K}}}, \text{for}~  i=1, 2$,
the PDF for the envelope of Simulators C and  D is given by
\begin{align}
 p_{Z}(z)=\left \{\begin{array}{ll} 4(1+K)z I_0\big(2\sqrt{1+K}z \big)K_0\big(2\sqrt{K}\big), & \mbox{for} \ z <\frac{\sqrt{K}}{\sqrt{1+K}}\\
4(1+K)z I_0\big(2\sqrt{K} \big)K_0\big(2\sqrt{1+K}z\big), & \mbox{for} \ z \geq\frac{\sqrt{K}}{\sqrt{1+K}} \label{y18}\\
\end{array} \right.
\end{align}
where $z=|h|$, $I_0$  is  the zero-order first kind modified Bessel function.

 Fig.~\ref{F2} shows the simulation of the PDF of envelopes for  Simulators A and  B.
  The plots indicate that distributions of envelopes of both simulators match the theoretical curves very well. For comparison, simulations of the PDF for the double-ring M-M mobile channel simulator~\cite{R16} is plotted. While the envelope of the double-ring simulator follows Rayleigh distribution, Simulators A and  B follow   cascaded  Rayleigh (worse than Rayleigh) distribution.

Fig.~\ref{F3} shows the simulation of the PDF of envelopes for  Simulators C and  D with different values of $K$. For the special case  $K=0$, Simulators C and  D becomes, respectively, Simulators A and  B up to a factor of $\sqrt{2}$.  This figure shows that the distribution of envelopes of both simulators match the theoretical curves in \eqref{y18} for all values of $K$.



\subsection{  Second-order statistics }

\textbf{Autocorrelation and cross-correlation of  Simulator B}. The autocorrelation of Simulator B is shown in Fig.~\ref{F7} using one and three trials. This figure suggests that Simulator B converges to the theoretical autocorrelation in one trial, and the difference between the simulation and the theoretical~\eqref{y21} is smaller in three trials. The cross-correlation is  shown in Fig.~\ref{F9} using one, three, and five trials. While simulated cross-correlation  using one trial converges to the theoretical~\eqref{y22} expression reasonably well, the simulated cross-correlation  converges to the theoretical expression more closely when  more trials are used.

\textbf{Autocorrelation and cross-correlation of  Simulator D}.
The autocorrelation of Simulator D  with different values of $K$ is shown in Fig.~\ref{F8}. These plots indicate that the simulated autocorrelation converges to the theoretical expression very well in one trial. It can also be observed that the LOS component dominates the   autocorrelation as $\tau$ increases. This is because the Bessel function approaches zero when  $\tau$ is large,  and  the autocorrelation thus approaches $K\cos\big(2 \pi f_3 \cos(\phi_3)\big)\big(2(1+K)\big)^{-1}$.
The cross-correlation for various values of $K$ is shown in Fig.~\ref{F10}. Again, the plots indicate that the simulation results converge to theoretical expressions in~\eqref{y27} and~\eqref{y30} in one trial.

Fig.~\ref{F11} plots the autocorrelation of the squared envelope for different values of $K$.  The simulated results converges to its theoretical in~\eqref{y32} in one trial. The plots suggests that the autocorrelation of the squared envelope approaches 1  when $\tau$ is large, regardless  of $K$. As $\tau\rightarrow\infty$, all Bessel terms in~\eqref{y32} become zero, which results in $R_{|h|^2|h|^2}(\tau)=\frac{K^2+2K+1} {(1+K)^2}=1$.

\textbf{Variance of the time-average correlations for Simulator  D.}
The variance of  time-average autocorrelation and cross-correlation for Simulator D with $K=0$ and $10$ are obtained by averaging $200$ samples  for each $\tau$.  As shown in Fig.~\ref{F17}, the variance time-average autocorrelation and cross-correlation are, respectively, in the magnitudes of $10^{-4}$ or $10^{-5}$.  Since the plot for $K=0$ reflects a variance of the time-average correlations for Simulator B up to a factor of $2$,  the small variance in the figure confirms that all proposed simulators have satisfactory convergence performance.

\subsection{ Higher-order statistics: LCR and AFD}

The level crossing rate  and the average fade duration  are two important statistical properties related to channel dynamics. At a specified level $R$, the LCR for a channel envelope is the rate (in crossings per second) at which the channel envelope crosses in the positive (or negative) direction~\cite{R15, R36, R38}. The AFD is the average time duration that the envelope remains below  the level $R$~\cite{R38, R36}. Approximations for LCR and AFD are plotted in~\cite{R40, R32} for the cascaded Rayleigh fading channels. Here, we present plots by numerical evaluation of the exact LCR and AFD functions.

\textbf{LCR for Simulator B}.
The LCR  for the cascaded Rayleigh channel without LOS has the following form~\cite{R34, R39}:
\begin{align}
L_{|g|}(R)=\big( 2\pi\sqrt{2}\big)^{\frac{1}{2}} R f_1 \int_{0}^{\infty}{x^{-2}}{\sqrt{a^2R^2+x^4}\ e^{-\frac{R^2+x^4}{\sqrt{2}x^2}}}dx \label{y34}
\end{align}
where $a=\frac{f_2}{f_1}$, and $R$ is the level. In the simulations of LCRs, the normalized LCRs and normalized signal level are used~\cite{R14, R15,R16, R38}. For simulator B, the normalized LCRs is $L_{|g|} f_1^{-1}$, and  normalized signal level is defined as  $\rho=\frac{R}{\sqrt{\mathcal{P}}}$, where $\sqrt{\mathcal{P}}$ is the root mean square (rms) envelope level for the channel, and  $\mathcal{P}=2$.  As shown  in Fig.~\ref{F12},  the simulated LCR matches well with the theoretical~\eqref{y34}. For comparison, we also simulated the LCR for a fixed-to-mobile channel~\cite{R18}. It can be observed that, for given signal levels, the LCR for Simulator B is more likely higher than that for fixed-to-mobile and double-ring models. This can be explained by the  higher dynamics of the statistical properties in cascaded Rayleigh fading  than a single Rayleigh fading distribution.

\textbf{LCR for Simulator D}.
%
%
The LCR for Simulator D is expressed as~\cite{R34, R39}:
\begin{align}
L_{|h|}(R)&=\big( 1+K\big)^{\frac{3}{4}}   R f_1\int_{0}^{\infty}\int_{-\pi}^{\pi}{x^{-2}}
\sqrt{a^2y_1({R}, {\vartheta})+x^4}\nn \\
&\times \biggl(y_2({R},{\vartheta},{x})
\Big(1+\text{erf}\Big(\frac{y_2({R},{\vartheta},{x})}
{\sqrt{2}}\Big)\Big) +\sqrt{\frac{2}{\pi}}e^{-\frac{y_2^2({R},{\vartheta},{x})}{2}}\biggl) e^{-\sqrt{1+K}(x^2+\frac{y_1({R}, {\vartheta})}{x^2})} d\vartheta dx \label{y35}
\end{align}
where $\text{erf}(\cdot)$ is the error function~\cite{R35}, and $y_1(R, \vartheta)$ and $y_2(R,\vartheta,x)$ are defined, respectively, as
\begin{align}
y_1(R, \vartheta)&=R^2+\frac{K}{1+K}-\frac{2R \sqrt{K}\cos(\vartheta)}{\sqrt{1+K}} \label{y36}\\
y_2(R,\vartheta,x)&=\frac{2 x f_3  \sqrt{K}\sin(\vartheta)}{f_1\sqrt{\sqrt{{1+K}}}\sqrt{a^2y_1(R,\vartheta)+x^4}}. \label{y37}
\end{align}
The LCRs of Simulator D with different values of $K $ are shown in Fig.~\ref{F13}. The  normalized signal level is $\rho={R}$, since the rms envelope level for Simulator D is one. While the plots indicate a good match between the simulated LCRs and the theoretical expression in~\eqref{y35}, it is also observed that the higher the $K$, the lower the observed LCRs.

\textbf{	AFD  for Simulator B}.
Using
\eqref{y34} and the CDF for $|g|$, the AFD for Simulator B can be obtained as~\cite{R36}:
\begin{align}
T_{|g|}(R)&=\frac{P_{Z}(z\leq R)}{L_{|g|}(R)}\!=\!\Big(1\!-\!\sqrt{2}RK_1\big(R\sqrt{2}\big)\Big)\Biggl(
f_1 R \sqrt{2\pi\sqrt{2}} \int_{0}^{\infty}{x^{-2}}{\sqrt{a^2R^2+x^4}e^{-\frac{R^2+x^4}{\sqrt{2}x^2}}}dx
\Biggl)^{-1} \label{y38}
\end{align}
In the simulations, normalized AFD~\cite{R14, R15, R16, R38,R18} is plotted. For Simulator B, the normalized AFD is $T_{|h|}f_1$.  The plots in Fig.~\ref{F14} shows agreement between the simulated  AFD  with the  theoretical~\eqref{y38}. For lower level of signals~($\rho<0$ dB), the AFD of Simulator B is longer than the AFDs of fixed-to-mobile and double-ring models; and for  higher level of signals~($\rho>0$ dB), the AFD of Simulator B is shorter than those  of fixed-to-mobile and double-ring models. This indicates that cascaded Rayleigh fading is more severe fading than a single Rayleigh fading.


\textbf{	AFD  for Simulator D}.
 The AFD can be calculated as
$
T_{|h|}(R)=\frac{P_{Z}(z\leq R)}{L_{|h|}(R)}$, where $L_{|h|}(R)$ is given in~\eqref{y35}, and the CDF of the envelope $z=|h|$ is given by
\begin{align}
 P_{Z}(Z \leq z)=\left \{\begin{array}{ll} 2\sqrt{1+K}z I_1\big(2\sqrt{1+K}z \big)K_0\big(2\sqrt{K}\big), & \mbox{for} \ z <\frac{\sqrt{K}}{\sqrt{1+K}}\\
1-2\sqrt{1+K}z I_0\big(2\sqrt{K} \big)K_1\big(2\sqrt{1+K}z\big), & \mbox{for} \ z \geq\frac{\sqrt{K}}{\sqrt{1+K}} \label{y19}\\
\end{array} \right.
\end{align}
where $I_0$  and $ I_1$ are the modified Bessel function first kind, zero-order and first-order, respectively.
Fig.~\ref{F15} shows the AFD for Simulator D. For lower level of signals~($\rho<0$ dB), the value of AFD decreases when the values of $K$ increase~(stronger component of LOS), and for  higher level of signals~($\rho>0$ dB), the value of AFD increases when the values of $K$ decrease~(weaker component of LOS).

\subsection{ 	Complexity analysis}
In this section, we examine the complexity of our proposed channel models. Since  simulators are sparse  for M-M cascaded Rayleigh fading channels in the literature, we compare Simulators A and B with the double-ring simulation model, one of the well-known simulators for M-M Rayleigh fading channels. It is worth to mention that double-ring and the proposed  simulators are for different categories of M-M channels, the comparison is solely for the illustration of complexity and performance of the proposed simulators.
The number of additions   required for Simulators A and B to generate one channel sample are, respectively, $2(P+Q)$ and $2(M+N)$, and $2MN$ additions are required for the double-ring model (same notations ($M$ and $N$) are used in~\cite{R16}). Besides additions, one multiplication is  required for both Simulators A and B, and no multiplication is required for the double-ring model. Table~\ref{Tab1} lists the averaged CPU time elapsed when generating one channel sample for Simulators A, B, and double-ring. A HP Compaq 8510p computer, Intel (R) Core (TM) 2 Duo CPU T7500 @ 2.20GHz, is used with $M=N=P=Q=8$ for all simulators. It can be seen that the elapsed time to generate one sample for  Simulators A and B is much shorter than that for the double-ring model. A relative CPU time is also listed, with $T_D$ being the referenced time for the double-ring model.

%

To investigate the performance of convergence, we examine the mean square error~(MSE) between simulated autocorrelation and the theoretical  for Simulators A, B, and double-ring model. These simulators have an identical theoretical autocorrelation expression~\eqref{y21}. The MSEs are obtained using one trial for different complexity levels. As shown in the first four rows in Table~\ref{Tab2}, if the same number of additions are used to generate one channel sample, then both Simulators A and B have much better convergence performance with significantly smaller MSE than the double-ring model, and  Simulator B provides the best convergence performance. If the same number of sinusoids are used (in this case, Simulators A and B have much less complexity),  the convergence performance of Simulators A and B is still comparable to that of the double-ring model.
Fig.~\ref{F16} plots the autocorrelations of Simulators A,  B, and double-ring model for the scenario listed in the second row of Table~\ref{Tab2}, whereby the number of additions to generate one channel sample is $200$ for all simulators.  The plots indicate that Simulator B has a faster convergence rate than the double-ring model.
\vspace{0.1cm}

\begin{table}[htbp]
\caption{CPU time elapsed for one channel sample}
\centering
\begin{tabular}{|p{1.8cm}|p{4.9cm}|p{1.3cm}|p{1.3cm}|}
    \hline Model   & Computations for one channel sample  &CPU time&  Relative CPU time  \\
                          \hline
     Simulator A &  $2 (Q+P)$ additions, $1$ multiplication & $3.8$s& 0.2 $T_D$ \\
    \hline
     Simulator B &  $2 (M+N)$ additions, $1$ multiplication & $4.3$s& 0.23 $T_D$ \\
    \hline
     Double-Ring & $2 M N$ additions & $18.8$s& $T_D$ \\
    \hline
\end{tabular}
\label{Tab1}
\end{table}

\begin{table}[htbp]
\caption{MSE and complexity}
\centering
\begin{tabular}{|p{3.1cm}|p{3.0cm}|p{2.5cm}|p{2.5cm}|}
    \hline Additions for One  & Simulator A  & Simulator B  & Double-Ring  \\
            Channel Sample    &             &               &   \\
    \hline
     \multirow{2}{*}{128} & $5.34\times10^{-4}$&$ 2.12\times10^{-4}$& $8.1\times10^{-3}$ \\
     & $\{Q=P=32\}$ & $\{M=N=32\}$  & $\{M=N=8\}$ \\
    \hline
      \multirow{2}{*}{200} & $3.19\times10^{-4}$& $1.80\times10^{-4}$& $4.69\times10^{-3}$ \\
     & $\{Q=P=50\}$ & $\{M=N=50\}$  & $\{M=N=10\}$ \\
    \hline
     \multirow{2}{*}{288} & $1.92\times10^{-4}$ & $1.64\times10^{-4}$&$ 2.91\times10^{-3}$ \\
     & $\{Q=P=72 \}$ & $\{M=N=72\}$  & $\{M=N=12\}$ \\
    \hline
     \multirow{2}{*}{392} & $1.78\times10^{-4}$& $1.28\times10^{-4}$& $2.40\times10^{-3}$ \\
     & $\{Q=P=98\}$ & $\{M=N=98\}$  & $\{M=N=14\}$ \\
    \hline
     $32$: Simulators A and B   & $7.3\times 10^{-3}$ & $ 2.8 \times 10^{-3}$  &  $8.1\times 10^{-3}$   \\
     $128$: Double-Ring & $\{Q=P=8\}$ & $\{M=N=8\}$ & $\{M=N=8\}$ \\
    \hline
\end{tabular}
\label{Tab2}
\end{table}

\section{Conclusion}
%
We have proposed statistical simulators  for  mobile-to-mobile channels, whereby the received signals experience cascaded Rayleigh fading with or without LOS. The simulators contain two individual summations and are therefore easy to implement with  lower complexity. Furthermore, the simulators provide faster convergence  to all the  desired statistical properties, including the pdf, autocorrelations, LCRs, and AFDs. Theoretical derivation of the time-averaged statistical properties and the corresponding variance are derived to confirm that the proposed simulators have  good convergence performance. Extensive Monte Carlo simulation results on various statistical properties and complexity analysis are provided to validate the proposed simulators.
While measurements and tests in various highly dense scattering environments confirm that the M-M channels may undergo cascaded Rayleigh fading~(more severe than single Rayleigh fading), our proposed simulators can be  used to simulate the  underlying channels  and  reveal the corresponding statical properties.

 %

\section*{acknowledgements}
 The authors are sincerely grateful  to Prof. M. Uysal for the valuable discussions and communications.


\appendix

\subsection{Proof of~\eqref{y21} and~\eqref{y22},  autocorrelation and cross-correlation for Simulator B}\label{A7}
From~\eqref{y5}, the real and imaginary parts of Simulator B are given, respectively, by
\begin{align}
g_c (t)&=\text{Re}\big(g(t)\big)=\sqrt{\frac{2}{NM}}
\sum_{n,m=1}^{N,M} \big( A_n(t)B_m(t)-C_n(t)D_m(t)\big)\label{n2}\\
g_s (t)&=\text{Im}\big(g(t)\big)=\sqrt{\frac{2}{NM}}
\sum_{n,m=1}^{N,M} \big( A_n(t)D_m(t)+C_n(t)B_m(t)\big) \label{n3}
\end{align}
\emph{For brevity of notations,  we }\emph{replace $t+\tau$ with $\varrho$  in the proceeding appendices}. The autocorrelation can be obtained by
\begin{align}
&\mathbb{E}\big[g_c(t) g_c(\varrho)\big]=\frac{2}{NM}\mathbb{E}\Biggl[
\sum_{n,m=1}^{N,M} \big( A_n(t)B_m(t)-C_n(t)D_m(t)\big)
 \sum_{k,j=1}^{N,M} \big( A_k(\varrho) B_j(\varrho)- C_k(\varrho) D_j(\varrho)\big) \Biggl]\nn \\ 
&~~~~~~~~~~~~~~= \frac{2}{NM}\mathbb{E}\Big[\sum_{n,k=1}^{N,N}  A_n(t)A_k(\varrho)\sum_{m,j=1}^{M,M}  B_m(t)B_j(\varrho)+ \sum_{n,k=1}^{N,N} C_n(t)C_k(\varrho)\sum_{m,j=1}^{M,M}  D_m(t)D_j(\varrho)\nn \\
&~~~~~~~~~~~~~~-\sum_{n,k=1}^{N,N}  A_n(t)C_k(\varrho)\sum_{m,j=1}^{M,M}  B_m(t)D_j(\varrho)-\sum_{n,k=1}^{N,N}  C_n(t)A_k(\varrho)\sum_{m,j=1}^{M,M}  D_m(t)B_j(\varrho)
\Big] \label{n4_d1}
\end{align}
 The cross-correlation is evaluated as
\begin{align}
&\mathbb{E}\big[g_c(t) g_s(\varrho)\big]=\mathbb{E}\Big[  \frac{2}{NM}
\sum_{n,m=1}^{N,M} \big( A_n(t)B_m(t)-C_n(t)D_m(t)\big)
\sum_{j,k=1}^{N,M} \big( A_j(\varrho)D_k(\varrho)+C_j(\varrho)B_k(\varrho)\big) \Big] 
\end{align}
Since   $\theta_n,  \Theta_n,  \Phi_m, \Psi_m \in [-\pi, \pi)$
are statistically independent and uniformly distributed for all $n$ and $m$, we have $\mathbb{E} [\sum_{n,k=1, n\neq k}^{N,N} A_n(t)A_k(\varrho)]\!=\!\mathbb{E} [\sum_{n,k=1, n\neq k}^{N,N}  C_n(t)C_k(\varrho)]=0$, and $ \mathbb{E} [\sum_{m,j=1, m\neq j}^{M,M}  B_m(t)B_j(\varrho)]= \mathbb{E} [\sum_{m,j=1, m\neq j}^{M,M}  D_m(t)D_j(\varrho)]=0$. It is easy to justify  $\mathbb{E}\big[g_c(t) g_c(\varrho)\big]=J_0(2\pi f_1 \tau) J_0(2\pi f_2 \tau)$ and $\mathbb{E}\big[g_c(t) g_s(\varrho)\big]=0$.

 Using~\cite{R35} (p. 420-421), the following identities are listed  for the convenience of the proceeding proof of the simulators' statistical properties.
 \begin{align}
 &\mathbb{E} [\sum_{n,k=1}^{N,N} A_n(t)A_k(\varrho)]=  \mathbb{E} [\sum_{n}^{N} A_n(t)A_n(\varrho)]=
 \!\frac{1}{2} \mathbb{E} [ \sum_{n=1}^{N}  \tilde{A}_n(\tau)]=  \!  \frac{N}{2} J_0(2\pi f_1 \tau) \label{factAA} \\
 &\mathbb{E} [\sum_{n,k=1}^{N,N} C_n(t)C_k(\varrho)]=  \mathbb{E} [\sum_{n}^{N} C_n(t)C_n(\varrho)]=
 \!\frac{1}{2} \mathbb{E} [\sum_{n=1}^{N}  \tilde{C}_n(\tau)]=  \!  \frac{N}{2} J_0(2\pi f_1 \tau) \label{factCC} \\
 &\!\!\!\sum_{m=1,j}^{M,M}  \! \mathbb{E}\![ B_m(t)B_j(\varrho)]\!=\!\!\sum_{m=1}^{M}  \mathbb{E}[ B_m(t)B_m(\varrho)]\!=\!\frac{1}{2} \mathbb{E} [\!\sum_{m=1}^{M} \tilde{B}_m(\tau)] =  \! \frac{M}{2} J_0(2\pi f_2 \tau\!) \label{factBB}\\
 &\!\!\!\sum_{m=1,j}^{M,M}  \! \mathbb{E}\![ D_m(t)D_j(\varrho)]\!=\!\!\sum_{m=1}^{M}  \mathbb{E}[ D_m(t)D_m(\varrho)]\!=\!\frac{1}{2} \!\sum_{m=1}^{M} \mathbb{E} [\tilde{D}_m(\tau) ]=  \! \frac{M}{2} J_0(2\pi f_2 \tau\!) \label{factDD}\\
 &\sum_{n,k=1}^{N,N}\mathbb{E}[A_n(t)C_k(\varrho)]=  \sum_{m,j=1}^{M,M}\mathbb{E}[B_m(t)D_j(\varrho)]=0. \label{fact_AC_BD}
\end{align}

%

\subsection{Proof of~\eqref{y27} and~\eqref{y30},  autocorrelation and cross-correlation for Simulator D}\label{A3}
The real and imaginary parts of Simulator D, respectively, are
\begin{align}
&h_c (t)=\text{Re}\big(h(t)\big)=\frac{1}{\sqrt{1+K}}\biggl(\sqrt{\frac{1}{2}}g_c(t)+\sqrt{K}
\cos\big( L(t)\big) \biggl), \label{y50} \\
&h_s (t)=\text{Im}\big(h(t)\big)=\frac{1}{\sqrt{1+K}}\biggl(\sqrt{\frac{1}{2}}
g_s(t)+\sqrt{K}\sin\big(L(t) \big) \biggl) \label{y51}
\end{align}
where
$L(t)=2\pi f_3 t \cos(\phi_3)+\phi_0$.
The autocorrelation is calculated as
\begin{align}
&\mathbb{E}\big[h_c(t) h_c(\varrho)\big]=\frac{1}{1+K}\mathbb{E}\Biggl[\biggl(\sqrt{\frac{1}{2}}g_c(t)+\sqrt{K}
\cos\big( L(t)\big) \biggl)\biggl(\sqrt{\frac{1}{2}}g_c(\varrho)+\sqrt{K}
\cos\big( L(\varrho)\big) \biggl) \Biggl] 
\end{align}
The cross-correlation of Simulator D can be obtained as
\begin{align}
&\mathbb{E}\big[h_c(t) h_s(t\!+\!\tau)\big]\!=\!\frac{1}{1+K}\mathbb{E}\Biggl[\biggl(\sqrt{\frac{1}{2}}g_c(t)\!+\!\sqrt{K}
\cos\big( L(t)\big) \biggl)\biggl(\sqrt{\frac{1}{2}}g_s(t\!+\!\tau)\!+\!\sqrt{K}
\sin\big( L(t\!+\!\tau)\big) \biggl) \Biggl]\label{y55}
\end{align}
Notice that the phase $\phi_0$ in $L(t)$ and $L(\varrho)$ is independent of other random variables in $g_c(t)$ and $g_s(t)$. Taking the expectation with respect to $\phi_0$ and using the results in the autocorrelation and cross-correlation for Simulator $B$, one can obtain the autocorrelation and cross-correlation as specified in~\eqref{y27} and~\eqref{y30}.

%
%

\subsection{Proof of~\eqref{ynew32}:  Squared envelope correlation for Simulator B}\label{A5_B}
The squared envelope correlation for Simulator B can be written as $R_{|g|^2|g|^2}(\tau)=
\mathbb{E}\big[g_c^2(t) g_c^2(\varrho)\big]+\mathbb{E}\big[g_c^2(t) g_s^2(\varrho)\big]+\mathbb{E}\big[g_s^2(t) g_c^2(\varrho)\big]+\mathbb{E}\big[g_s^2(t) g_s^2(\varrho)\big]$.
The first term is expressed as
\begin{align}
\mathbb{E}&\big[g_c^2(t) g_c^2(\varrho)\big]=\frac{4}{N^2M^2}
\mathbb{E}\bigg[
\sum_{n,m=1}^{N,M} \big(A_n(t)B_m(t) - C_n(t)D_m(t)\big)
\sum_{u,p=1}^{N,M} \big(A_u(t)B_p(t)- C_u(t)D_p(t)\big)  \nn\\
&~~~~~~~~~~~~~~~~~~~~~~\times \sum_{q,r=1}^{N,M}\big( A_q(\varrho)B_r(\varrho) -C_q(\varrho) D_r(\varrho)\big)
\sum_{s,j=1}^{N,M} \big( A_s(\varrho) B_j(\varrho) - C_s(\varrho)D_j(\varrho)\big)\bigg].\label{appe_squared_gg_1}
\end{align}
Expanding~\eqref{appe_squared_gg_1} and taking expectation with respect to the phases, we obtain
$\mathbb{E}\big[g_c^2(t) g_c^2(\varrho)\big]=\frac{4}{N^2M^2}(\Upsilon_A \Upsilon_B+\Upsilon_C \Upsilon_D+ \frac{M^2N^2}{8}+ 4 \kappa)$, where $\Upsilon_X= \mathbb{E}\big[\sum_{n}^{N} X_n(t)  \sum_{u}^{N} X_u(t) \sum_{q}^{N} X_q(\varrho) \sum_{s}^{N} X_s(\varrho)\big]$, where $X=\{A,B,C,D\}$, and
\begin{align}
\kappa&=\mathbb{E}\big[ \sum_{n,s}^{N,N} A_n(t) A_s(\varrho) \sum_{u,q}^{N,N} C_u(t) C_q(\varrho)
\sum_{m,j}^{M,M}  B_m(t)B_j(\varrho) \sum_{p,r=1}^{M,M}D_p(t) D_r(\varrho)]\label{AAAA_0}
\end{align}
Similarly, we have $\mathbb{E}\big[g_c^2(t) g_s^2(\varrho)\big]=\mathbb{E}\big[g_s^2(t) g_c^2(\varrho)\big] = \frac{4}{N^2M^2}\big( \frac{M^2}{4}( \Upsilon_A+\Upsilon_C)+\frac{N^2}{4}(\Upsilon_B+\Upsilon_D)-4 \kappa\big)$, and $\mathbb{E}\big[g_s^2(t) g_s^2(\varrho)\big]=\frac{4}{N^2M^2}(\Upsilon_A \Upsilon_D+\Upsilon_C \Upsilon_B+ \frac{M^2N^2}{8}+ 4 \kappa)$. 
The term  $\Upsilon_A$ is evaluated as
\begin{align}
\Upsilon_A
&= \mathbb{E}\Big[  \sum_p^N A^2_p(t)\sum_j^N A^2_j(\varrho)\Big]+\mathbb{E}\Big[  \sum_{n,u, n \neq u}^{N,N} A_n(t)A_u(t)\sum_{q,s, q \neq s}^{N,N} A_q(\varrho)A_s(\varrho) \Big]\label{AAAA_1}
\end{align}
The first term in~\eqref{AAAA_1} is obtained as
\begin{align}
&\mathbb{E}\big[  \sum_p^N A^2_p(t)\sum_j^N A^2_j(\varrho)\big]
=\frac{2N^2+NJ_0(4 \pi f_1 \tau)}{8}.\label{AAAA_2}
\end{align}
The second term in~\eqref{AAAA_1} contains the following seven cases:
\begin{itemize}
\item Case 1. $n \neq u, q\neq s, n=q, u \neq s$;
\item Case 2. $n \neq u, q\neq s, n=s, q \neq u$;
\item Case 3. $n \neq u, q\neq s, u=q, n \neq s$;
\item Case 4. $n \neq u, q\neq s, u=s, n \neq q$;
\item Case 5. $n \neq u, q\neq s, n \neq q, u \neq s$;
\item Case 6. $n \neq u, q\neq s, n=q, u=s$;
\item Case 7. $n \neq u, q\neq s, n=s, u=q$.
\end{itemize}
 The value of $ \mathbb{E}\big[ \sum_{n}^{N} A_n(t)A_n(\varrho)\sum_{s, s\neq n}^{N} A_s(t)A_s(\varrho)\big]$ is zero for Cases 1 to 5 and identical for  Cases 6 and 7, and  $ \mathbb{E}\big[ \sum_{n}^{N} A_n(t)A_n(\varrho)\sum_{s, s\neq n}^{N} A_s(t)A_s(\varrho)\big]_{\text{case 6 or 7}}
=\frac{N^2J^2_0(2 \pi f_1 \tau) - \sum_{n}^{N} \big( \mathbb{E}[\tilde{A}_n(\tau)]\big)^2}{4}
$. 
Therefore, we have
$
\Upsilon_A= \frac{2N^2+NJ_0(4 \pi f_1 \tau)+ 4N^2J^2_0(2 \pi f_1 \tau)}{8}- \frac{ \sum_{n}^{N} \big( \mathbb{E}[\tilde{A}_n(\tau) ]\big)^2}{2}
$. In fact, the value of $\Upsilon_A$ can also be obtained following the steps in (48) Appendix I~\cite{R42}.

Similarly, we have
$
\Upsilon_C= \frac{2N^2+NJ_0(4 \pi f_1 \tau)+ 4N^2J^2_0(2 \pi f_1 \tau)}{8} -\frac{\sum_{n}^{N} \big( \mathbb{E}[\tilde{C}_n(\tau) ]\big)^2}{2}
$. It is straightforward to justify $\mathbb{E}\big[\tilde {C}_n(\tau)]=\mathbb{E}\big[\tilde {C}_{N-n}(\tau)], n=1, \cdots, N $ by the following steps:
\begin{align}
\mathbb{E}\big[\tilde {C}_n(\tau)]&= \int_{-\pi}^{\pi}\cos \big( 2 \pi f_1 \tau \cos (\frac{2n\pi - \pi +\psi}{4N})\big)\frac{1}{2\pi} d\psi \nn \\
&= \int_{\frac{(n-1)\pi}{2N}}^{\frac{n \pi}{2N}}\cos \big( 2 \pi f_1 \tau \cos (\theta)\big) \frac{4N}{2\pi} d\psi\nn \\
& = \int_{\frac{(N-n-1)\pi}{2N}}^{\frac{(N-n) \pi}{2N}}\cos \big( 2 \pi f_1 \tau \sin (\varphi)\big) \frac{4N}{2\pi} d\varphi = \mathbb{E}\big[\tilde {C}_{N-n}(\tau)]
\end{align}
Denote $ \xi(f_1,\tau)= \sum_{n}^{N} \Big( \mathbb{E}\big[\tilde{A}_n(\tau) \big]\Big)^2=\sum_{n}^{N} \Big( \mathbb{E}\big[\tilde{C}_n(\tau) \big]\Big)^2$. Then, we have
 \begin{align}
 \Upsilon_A&= \Upsilon_C= \frac{2N^2+NJ_0(4 \pi f_1 \tau)+ 4N^2J^2_0(2 \pi f_1 \tau)}{8} -\frac{\xi(f_1,\tau)}{2} \\
 \Upsilon_B&= \Upsilon_D= \frac{2M^2+MJ_0(4 \pi f_2 \tau)+4 M^2J^2_0(2 \pi f_2 \tau)}{8}-\frac{ \xi(f_2,\tau)}{2},
 \end{align}
 where
 $\xi(f_2,\tau)=\sum_{m}^{M} \Big( \mathbb{E}\big[\tilde{B}_m(\tau) \big]\Big)^2=\sum_{m}^{M} \Big( \mathbb{E}\big[\tilde{D}_n(\tau) \big]\Big)^2$.
%
Inserting the results for  $ \mathbb{E}\big[g_c^2(t) g_c^2(\varrho)\big], \mathbb{E}\big[g_c^2(t) g_s^2(\varrho)\big], \mathbb{E}\big[g_s^2(t) g_c^2(\varrho)\big]$, and $\mathbb{E}\big[g_s^2(t) g_s^2(\varrho)\big] $ into $R_{|g|^2|g|^2}(\tau)$, we obtain
\begin{align}
R_{|g|^2|g|^2}(\tau)
&= \frac{4}{N^2M^2}\bigg(\Upsilon_A +\Upsilon_C+ \frac{N^2}{2}\bigg)\bigg(\Upsilon_B+ \Upsilon_D+\frac{M^2}{2}\bigg) \nn \\
\!=\!\frac{4}{N^2M^2}\Big(\!&N^2\!\!+\! N^2J^2_0(2 \pi f_1 \tau) \!+\! \frac{NJ_0(4 \pi f_1 \tau)}{4}  \!-\!\xi(f_1, \tau) \Big )\!
\Big(\!M^2\!\!+\! M^2J^2_0(2 \pi f_2 \tau)\!+\!\frac{MJ_0(4 \pi f_2 \tau)}{4}\!-\!\xi(f_2,\tau) \Big).\label{appe_auto_ggsquare_final}
\end{align}
Expanding~\eqref{appe_auto_ggsquare_final} yields~\eqref{ynew32}.

\subsection{Proof of~\eqref{y32}: Squared envelope correlation for Simulator D}\label{A5_D}
The squared envelope correlation for Simulator D contains four terms as indicated in~\eqref{correlation_square_hh}.
The first term is evaluated as
\begin{align}
&\mathbb{E}\big[h_c^2(t) h_c^2(\varrho)\big]=\frac{1}{(1+K)^2}\mathbb{E}\Biggl[\biggl(\sqrt{\frac{1}{2}}g_c(t)+\sqrt{K}
\cos\big( L(t)\big) \biggl)^2\biggl(\sqrt{\frac{1}{2}}g_c(\varrho)+\sqrt{K}
\cos\big( L(\varrho)\big) \biggl)^2\biggl] \nn\\
&~~~~~~~~~~~~~~~~~=\frac{1}{(1+K)^2}\mathbb{E}\Biggl[\biggl({\frac{1}{2}}g^2_c(t)+ \sqrt{2K} g_c(t)\cos\big( L(t)\big)
+ K \cos^2 \big(L(t)\big) \biggl)\nn \\
&~~~~~~~~~~~~~~~~~~~~~~~~~~~~~~~~~\times
\biggl({\frac{1}{2}}g^2_c(\varrho)+ \sqrt{2K} g_c(\varrho)\cos\big( L(\varrho)\big)+ K \cos^2 \big(L(\varrho)\big) \biggl)\biggl]\label{appe_squared_hh_1}
\end{align}
Note that $\mathbb{E}[g_c(t)]=\mathbb{E}[g_c(\varrho)]=0$, $\mathbb{E}[\cos\big( L(\ell t)\big)]=\mathbb{E}[\cos\big( \ell L(\varrho)\big)]=0,\ell=1,2$, and $\mathbb{E}[g^2_c(t)] =\mathbb{E}[g^2_c(\varrho)]=R_{g_cg_c}(0)=1$. We have
\begin{align}
&\mathbb{E}\big[h_c^2(t) h_c^2(\varrho)\big]=\frac{1}{(1+K)^2}\Biggl(\frac{1}{4}\mathbb{E}\big[g_c^2(t) g_c^2(\varrho)\big] + {2K}R_{g_c g_c}(\tau)\mathbb{E}\big[ \cos \big(L(t)\big) \cos \big(L(\varrho)\big)\big]\nn \\
&~~~~~~~~~~~~~~~~~~~~~~~~~~+\frac{K}{2} \mathbb{E}\big[ \cos^2 \big(L(t)\big) +\cos^2 \big(L(\varrho)\big) \big]
+K^2 \mathbb{E}\big[ \cos^2 \big(L(t)\big) \cos^2 \big(L(\varrho)\big) \big]
\Biggl)\label{appe_squared_hh_2}
\end{align}
The autocorrelation of the squared quadrature component $\mathbb{E}\big[h_s^2(t) h_s^2(\varrho)\big]$ can be evaluated following similar steps in obtaining
$\mathbb{E}\big[h_c^2(t) h_c^2(\varrho)\big]$.  Using identities
 $\mathbb{E}\big[ \cos \big(L(t)\big) \cos \big(L(\varrho)\big)\big]= \frac{1}{2}\cos\big(2\pi f_3 \tau \cos(\phi_3)\big), \mathbb{E}\big[ \cos^2 \big(L(t)\big)\big]=\mathbb{E}\big[\cos^2 \big(L(\varrho)\big) \big]=\frac{1}{2} $, and $\mathbb{E}\big[ \cos^2 \big(L(t)\big) \cos^2 \big(L(\varrho)\big) \big]= \frac{1}{4}+
\frac{1}{8}\cos\big(4\pi f_3 \tau \cos(\phi_3)\big)$, we summarize the auto- and cross-correlations, respectively, as  
\begin{align}
&\mathbb{E}\big[h_x^2(t) h_x^2(\varrho)\big]\!=\!\frac{2 \mathbb{E}\big[g_x^2(t) g_x^2(\varrho)\big]
 \!+\!8K R_{g_x g_x}(\tau) \cos\big(2\pi f_3 \tau \cos(\phi_3)\big)
\!+\!{4K\!+\! 2K^2\!\!+\!K^2\!\! \cos\big(4\pi f_3 \tau \cos(\phi_3)\big)}}{8(1+K)^2} \label{appe_squared_hh_3}\\
&\mathbb{E}\big[h_c^2(t) h_s^2(\varrho)\big]= \mathbb{E}\big[h_s^2(t) h_c^2(\varrho)\big]=\frac{2\mathbb{E}\big[g_c^2(t) g_s^2(\varrho)\big]+\!{4K\!+\! 2K^2\!-\!K^2 \cos\big(4\pi f_3 \tau \cos(\phi_3)\big)}}{8(1+K)^2} \label{appe_squared_hh_4}
\end{align}
where $x=\{c,s\}$.
%
 Inserting~\eqref{appe_squared_hh_3} and~\eqref{appe_squared_hh_4} into~\eqref{correlation_square_hh} yields~\eqref{y32}.


\subsection{Proof of~\eqref{r4}: Variance of  time-average correlations for Simulator B}\label{A6_B}
The variance of the time-average autocorrelation of the real part for Simulator B is $\text{Var}[\tilde{R}_{g_c g_c}(\tau) ] =\mathbb{E}[ \tilde{R}_{g_c g_c}^2(\tau)]-({R}_{g_c g_c}(\tau))^2$.
 The first term is evaluated as
\begin{align}
\mathbb{E}&[ \tilde{R}_{g_c g_c}^2(\tau)]\!= \frac{1}{4M^2N^2 }
\mathbb{E}\Big[\sum_{n,m=1}^{N,M} \!\!\!\!\big(\tilde{A}_n(\tau) \tilde{B}_m(\tau) \!+\!\tilde{C}_n(\tau) \tilde{D}_m(\tau)\big)
\sum_{p,q=1}^{N,M} \!\!\!\big(\tilde{A}_p(\tau) \tilde{B}_q(\tau) \!+\!\tilde{C}_p(\tau) \tilde{D}_q(\tau)\big)\Big ]\nn \\
&~~~~~~~~~~~= \frac{1}{4N^2M^2} \Biggl(  \mathbb{E}\biggl[\sum_{n,q=1}^{N,N}\!\!\tilde{A}_n(\tau)\tilde{A}_q(\tau)\!\!\sum_{m,p=1}^{M,M}\!\!\tilde{B}_m(\tau)\tilde{B}_p(\tau)  \biggl]
\!+2 \mathbb{E}\biggl[\sum_{n,q=1}^{N,N}\!\!\tilde{A}_n(\tau)\tilde{C}_q(\tau)\!\!\sum_{m,p=1}^{M,M}\!\!\tilde{B}_m(\tau)\tilde{D}_p(\tau)  \biggl]\nn\\
&~~~~~~~~~~~~~~~~~~~~~~~~
+\mathbb{E}\biggl[\sum_{n,q=1}^{N,N}\tilde{C}_n(\tau)\tilde{C}_q(\tau)\sum_{m,p=1}^{M,M}\tilde{D}_m(\tau)\tilde{D}_p(\tau \biggl]
\Biggl)\label{appe_var time-average g_1}
\end{align}
We have the following identities:
\begin{align}
\mathbb{E}\big[\sum_{n,q=1}^{N,N}\tilde{A}_n(\tau)\tilde{A}_q(\tau)\big] &=N^2 J^2_0(2 \pi f_1 \tau)+ \mathbb{E}\big[\sum_{n=1}^N \tilde{A}_n(\tau) \tilde{A}_n(\tau) \big]- \sum_{n=1}^N  \mathbb{E}\big[ \tilde{A}_n(\tau) \big]\mathbb{E}\big[ \tilde{A}_n(\tau) \big]~~~~~~~~~~~~~~~\nn \\
&= N^2 J^2_0(2 \pi f_1 \tau)+ \frac{N}{2}+\frac{N J_0(4 \pi f_1 \tau)}{2}- {\xi(f_1,\tau)}\label{appe_AA_tilde}
\end{align}
\begin{align}
 & \mathbb{E}\big[\sum_{n,q=1}^{N,N}\tilde{C}_n(\tau)\tilde{C}_q(\tau)\big] = \mathbb{E}\big[\sum_{n,q=1}^{N,N}\tilde{A}_n(\tau)\tilde{A}_q(\tau)\big]\label{appe_CC_tilde}\\
  & \mathbb{E}\big[\sum_{m,p=1}^{M,M}\tilde{B}_m(\tau)\tilde{B}_p(\tau) \big] \!=\!
    \mathbb{E}\big[\sum_{m,p=1}^{M,M}\tilde{D}_m(\tau)\tilde{D}_p(\tau) \big]\! =\! M^2 J^2_0(2 \pi f_2 \tau)\!+\! \frac{M}{2}\!+\!\frac{M J_0(4 \pi f_2 \tau)}{2}- {\xi(f_2,\tau)}\label{appe_BB_DD_tilde}\\
      & \mathbb{E}\big[\sum_{n,q=1}^{N,N}\tilde{A}_n(\tau)\tilde{C}_q(\tau) \big] = N^2 J^2_0(2 \pi f_1 \tau)+V_{\tilde{A}\tilde{C}}\label{appe_AC_tilde} \\
       & \mathbb{E}\big[\sum_{m,p=1}^{M,M}\tilde{B}_m(\tau)\tilde{D}_p(\tau) \big] = M^2 J^2_0(2 \pi f_2 \tau)+V_{\tilde{B}\tilde{D}}\label{appe_BD_tilde}
      \end{align}
 where 
$V_{\tilde{A}\tilde{C}}= \sum_{n=1}^N \big( \mathbb{E}[ \tilde{A}_n(\tau) \tilde{C}_n(\tau)]-\mathbb{E}[ \tilde{A}_n(\tau)]\mathbb{E}[ \tilde{C}_n(\tau)] \big)$,
$V_{\tilde{B}\tilde{D}}= \sum_{m=1}^M \big( \mathbb{E}[ \tilde{B}_m(\tau) \tilde{D}_m(\tau)]-\mathbb{E}[ \tilde{B}_m(\tau)]\mathbb{E}[ \tilde{D}_m(\tau)] \big)$. Inserting~\eqref{appe_AA_tilde} to~\eqref{appe_BD_tilde}
into~\eqref{appe_var time-average g_1}, we obtain
\begin{align}
\mathbb{E}&[ \tilde{R}_{g_c g_c}^2(\tau)]\!= \frac{1}{4M^2N^2 } \Big(
2 \big(N^2 J^2_0(2 \pi f_1 \tau)+ \frac{N}{2}+\frac{N J_0(4 \pi f_1 \tau)}{2}- {\xi(f_1,\tau)}\big)\nn \\
&\times \!\! \big( M^2 J^2_0(2 \pi f_2 \tau)\!+\! \frac{M}{2}\!+\!\frac{M J_0(4 \pi f_2 \tau)}{2}\!- {\xi(f_2,\tau)}\big)
\!+\! 2 \big(N^2 J^2_0(2 \pi f_1 \tau)\!+\!V_{\tilde{A}\tilde{C}}\big)\big(M^2 J^2_0(2 \pi f_2 \tau)\!+\!V_{\tilde{B}\tilde{D}}\big)
\Big)\nn \\
&= \frac{1}{8M^2N^2 } \Big( \big(2N^2 J^2_0(2 \pi f_1 \tau)\!+\! N\!+\!{N J_0(4 \pi f_1 \tau)}\!-\!2~ {\xi(f_1,\tau)}\big)\nn \\
&\times \!\!\big( 2M^2 J^2_0(2 \pi f_2 \tau)\!+\!\! M\!+\!\!{M J_0(4 \pi f_2 \tau)}\!-\!2~{\xi(f_2,\tau)}\big)
\!+\! 4 \big(N^2 J^2_0(2 \pi f_1 \tau)\!+\!V_{\tilde{A}\tilde{C}}\big)\big(M^2 J^2_0(2 \pi f_2 \tau)\!+\!V_{\tilde{B}\tilde{D}}\big)\Big)
\end{align}
Recall  $\big(\tilde{R}_{g_c g_c} (\tau)\big)^2=J^2_0 (2\pi f_1 \tau)J^2_0 (2\pi f_2 \tau)$. We have
\begin{align}
&\text{Var}[\tilde{R}_{g_c g_c}(\tau) ] 
= \frac{1}{8M^2N^2 } \Big(  2 M^2 J^2_0(2 \pi f_2 \tau) \big( N+{N J_0(4 \pi f_1 \tau)}- 2~ {\xi(f_1,\tau)}+2V_{\tilde{A}\tilde{C}}  \big)\nn \\
&~~~~~~~~~~~~~~~~~+2 N^2 J^2_0(2 \pi f_1 \tau)\big( M+{M J_0(4 \pi f_2 \tau)}-2~ \xi(f_2,\tau) +2V_{\tilde{B}\tilde{D}}\big)\nn \\
&~~~~~~~~~~~~~~~~~+\big( N\!+\!{N J_0(4 \pi f_1 \tau)}-2~ \xi(f_1,\tau) \big) \big( M\!+\!{M J_0(4 \pi f_2 \tau)}-2~\xi(f_2,\tau)\big)
\!+\!4V_{\tilde{A}\tilde{C}}V_{\tilde{B}\tilde{D}}\Big) \label{appe_r4}
\end{align}
Reorganizing~\eqref{appe_r4}, one can obtain~\eqref{r4}.

The variance of time-average autocorrelation of the imaginary part of Simulator B is $\text{Var}[\tilde{R}_{g_s g_s}(\tau) ] =\mathbb{E}[ \tilde{R}_{g_s g_s}^2(\tau)]-({R}_{g_s g_s}(\tau))^2$.
 The first term is evaluated as
\begin{align}
\mathbb{E}&[ \tilde{R}_{g_s g_s}^2(\tau)]\!= \frac{1}{4M^2N^2 }
\mathbb{E}\Big[\sum_{n,m=1}^{N,M} \!\!\!\!\big(\tilde{A}_n(\tau) \tilde{D}_m(\tau) \!+\!\tilde{C}_n(\tau) \tilde{B}_m(\tau)\big)
\sum_{p,q=1}^{N,M} \!\!\!\big(\tilde{A}_p(\tau) \tilde{D}_q(\tau) \!+\!\tilde{C}_p(\tau) \tilde{B}_q(\tau)\big)\Big ]\nn \\
&~~~~~~~~~~~= \frac{1}{4N^2M^2} \Biggl(  \mathbb{E}\biggl[\sum_{n,q=1}^{N,N}\!\!\!\tilde{A}_n(\tau)\tilde{A}_q(\tau)\!\!\sum_{m,p=1}^{M,M}\!\!\!\tilde{D}_m(\tau)\tilde{D}_p(\tau)  \biggl]
+2 \mathbb{E}\biggl[\sum_{n,q=1}^{N,N}\!\!\!\tilde{A}_n(\tau)\tilde{C}_q(\tau)\!\!\sum_{m,p=1}^{M,M}\!\!\!\tilde{B}_m(\tau)\tilde{D}_p(\tau)  \biggl]\nn\\
&~~~~~~~~~~~~~~~~~~~~~~~~~+\mathbb{E}\biggl[\sum_{n,q=1}^{N,N}\tilde{C}_n(\tau)\tilde{C}_q(\tau)\sum_{m,p=1}^{M,M}\tilde{B}_m(\tau)\tilde{B}_p(\tau)  \biggl]\Biggl)\label{appe_var time-average g_2}
\end{align}
We have $\text{Var}[\tilde{R}_{g_s g_s}(\tau) ] =\text{Var}[\tilde{R}_{g_c g_c}(\tau) ] $.

\subsection{Proof of~\eqref{r5}: Variance of time-average correlations for Simulator D}\label{A6_D}
We present the derivation of the variance of the time-average autocorrelation of the in-phase component in~\eqref{r5}, while the same steps can be applied to obtain  the variance of the time-average autocorrelation of the quadrature component.
The variance of the time-average autocorrelation of the in-phase component for simulator D is $\text{Var}\big[\tilde{R}_{{h_c}h_c}(\tau) \big] =\mathbb{E}[\tilde{R}_{{h_c}h_c}^2(\tau)]- \big( {R}_{{h_c}h_c}(\tau)   \big)^2$. The first term is evaluated as
\begin{align}
\mathbb{E}[\tilde{R}_{{h_c}h_c}^2(\tau)]&
=\mathbb{E}\biggl[\biggl(\frac{\tilde{R}_{{g_c}g_c}(\tau) +K \cos\big( 2\pi f_3 \tau \cos(\phi_3) \big) }
{2(1+K)} \biggl)^2\biggl]\nn\\
=\frac{1}{4(1+K)^2}& \Big( \mathbb{E}\big[   \tilde{R}_{{g_c}g_c}^2(\tau)\big]+ 2K\cos\big(2 \pi f_3 \tau \cos(\phi_3) \big)\mathbb{E}\big[ \tilde{R}_{{g_c}g_c}(\tau)\big]+\big( K\cos(2\pi f_3\tau \cos(\phi_3)) \big)^2\Big)\label{var time-average 2}
\end{align}
Recall $\mathbb{E}[\tilde{R}_{g_c g_c} (\tau)]=J_0 (2\pi f_1 \tau)J_0 (2\pi f_2 \tau)$. Using $\mathbb{E}\big[   \tilde{R}_{{g_c}g_c}^2(\tau)\big]$   obtained in Appendix~\ref{A6_B} and ${R}_{{h_c}h_c}(\tau)  $ from Appendix~\ref{A3}, it is easy to obtain~\eqref{r5}.

\subsection{Proof for PDF of envelope for Simulators A and B}\label{A1}

The pdf and phase distribution of Simulators A and B are readily available in the literature~\cite{R33}.
Denote $g=g_1g_2=g_c+jg_s$, $g_1=x_1+jx_2$, $g_2=y_1+jy_2$, $x_1, x_2\sim \mathcal{N}(0, \sigma_1^2)$, and $y_1, y_2\sim \mathcal{N}(0, \sigma_2^2)$ are i.i.d. zero mean Gaussian normally distributed. The joint PDF for the real and imaginary parts of $g$ is given as $
p_{g_cg_s}(g_c,g_s)=\frac{1}{2\pi\sigma_1^2\sigma_2^2 } K_0\biggl(\frac{\sqrt{g_c^2+g_s^2}}{\sigma_1\sigma_2}\biggl) $.
%
To find the pdf of the envelope, we transform  the Cartesian coordinates $(g_c, g_s)$ to polar coordinates
$(z, \eta)$, where $ z=\sqrt{g_c^2+g_s^2}$, and
$\eta=\arctan\Big(\frac{g_s(t)}{g_c(t)}\Big)$. The resulted transformation Jacobian is
${1}/{z}$. The joint pdf of the envelope and phase is given by
$
p_{z,\eta}(z,\eta)=z p_{g_cg_s}(g_c,g_s)=\frac{z}{2\pi\sigma_1^2\sigma_2^2 } K_0\biggl(\frac{z}{\sigma_1\sigma_2}\biggl) $.
The PDF of the envelope is obtained as
$
p_{z}(z)=
\frac{z}{\sigma_1^2\sigma_2^2 } K_0\biggl(\frac{z}{\sigma_1\sigma_2}\biggl) $.
Recognizing   
 $\sigma_1=\sigma_2={1}/{\sqrt{\sqrt{2}}}$ for Simulators A and B, one can obtain the PDF of their envelopes.

\vspace{-0.2cm}
\nocite{Scharf91,nocedal_book}
\begin{small}
\bibliographystyle{IEEEbib}
\bibliography{strings,refs,manuals,mybib1}
\renewcommand\refname{References}

\end{small}

\end{document}